
\baselineskip=\normalbaselineskip \multiply\baselineskip by 2
\magnification=1200
\def\C#1#2#3{C^{#1_0}_{#2_0 #3_0}}
\def\D#1#2#3#4{D^{{#1}_{#4}} _{#2_0 #3_{#4}}}

\def\Z#1#2#3#4{Z^{#1_#3}_{#2_#4}}
\def\x#1#2{\xi^{#1} _{#2_0}}
\def\bx#1#2{\xi^{#1} _{[#2_0}}

\def\vx#1{{\vec \xi}_{{#1}_0}}
\def\vXi#1{{\vec \Xi}_{{#1}_0}}
\def\vX1#1{{\vec \Xi}_{{#1}_1}}
\def\G#1{G_{#1_0}}
\def\hG#1{{\hat G}_{#1_0}}
\def\M#1#2#3#4#5#6{M^{#1_{#5}}_{#2_0 #3_0 #4_{#6}}}
\def\B#1#2#3#4#5{B^{#1_{#4}}_{{#2}_1#3_{#5}}}
\def\e#1#2{\eta^{#1_{#2}}}
\def\P#1#2{{\cal P}_{#1_{#2}}}
\def\hP#1#2{{\hat{\cal P}}_{#1_{#2}}}
\def\M#1#2#3#4#5#6{M^{#1_{#5}}_{#2_0 #3_0 #4_{#6}}}
\def\hO{\hat\Omega}
\def\A#1#2#3{A_{#1_#3}\ ^{#2_#3}}
\def\Ai#1#2#3{\left(A^{-1}\right)_{#1_#3}\ ^{#2_#3}}
\rightline{ULB-PMIF/92-07}
\rightline{GTCRG/92-04}
\vskip3truecm
\centerline{\bf ON THE QUANTIZATION OF REDUCIBLE GAUGE SYSTEMS}
\vskip1pc
\centerline{Rafael Ferraro$^\dagger$, Marc Henneaux$^{\dagger\dagger}$ and
Marcel  Puchin$^{\dagger\dagger\dagger}$}
\vskip1pc
\centerline{\it Facult\'e des Sciences, Universit\'e Libre de Bruxelles,
Campus Plaine,}
\centerline{\it C.P.231, B-1050 Bruxelles, Belgium.}
\vskip3pc
{\bf Abstract:}  Reducible  gauge theories with constraints linear in the
momenta are quantized.    The  equivalence  of  the  reduced  phase space
quantization, Dirac quantization and  BRST  quantization  is established.
The ghosts of ghosts are  found to play a crucial role in the equivalence
proof.
\vfill
\hrule
{\baselineskip=\normalbaselineskip \multiply\baselineskip by 0
\noindent$^\dagger$ On leave of absence from
{\it Instituto de Astronom\'{\i}a y
F\'{\i}sica del Espacio, C.C.167, Sucursal 28, 1428 Buenos Aires, Argentina,}
and {\it Facultad de Ciencias Exactas y Naturales, Departamento de
F\'{\i}sica, Universidad de Buenos Aires, Ciudad Universitaria, Pab.I, 1428
Buenos Aires, Argentina.}
\vskip1pc
\noindent$^{\dagger\dagger}$ Also at {\it Centro de Estudios Cient\'{\i}ficos
de Santiago, Casilla 16443, Santiago 9, Chile.}
\vskip1pc
\noindent$^{\dagger\dagger\dagger}$On leave of absence from {\it Universitatea
din  Craiova,  Str.A.I.Cuza  13,  Facultatea  de  Fizica,  1100  Craiova,
Romania.} \vfill }
\eject
\centerline{\bf 1. INTRODUCTION}
\vskip2pc
Gauge theories can be quantized according to at least three different
methods:
\smallskip
\noindent (i) The reduced phase space method quantizes only the
gauge invariant functions and is for that reason physically
quite appealing. However, it is often not tractable because
it requires the explicit finding of a complet set of gauge
invariant functions.
\smallskip
\noindent (ii) The Dirac method realizes all the dynamical variables
(gauge invariant and non gauge invariant ones) as operators
in some linear space of states, and selects the physical states by
means of a subsidiary condition.
\smallskip
\noindent (iii) The BRST method increases further the redundancy in
the description of the system by introducing ghosts. The
physical states are again selected by means of a subsidiary
condition.
\smallskip
It is easy to check that the three different approaches to
the quantization of gauge systems are equivalent in the case
of simple constraints (see for instance Ref.$[1]$).
The question of their equivalence for arbitrary
systems is  more subtle and has attracted recently a considerable amount of
interest $^{1-14}$. Because the problem of ``quantization'' is inherently
ambiguous (many different quantum systems possess the same
$\hbar\rightarrow  0$  limit), the question of  equivalence  is  actually
ill-defined  in  the  absence  of  a  definite   choice  of  quantization
prescriptions.    For  this reason, a conclusive analysis  should  either
exhibit  correspondence  rules  that  insure  equivalence  of  the  three
quantization methods, or prove the inexistence of such rules.

The previous works on the equivalence question are all
devoted to  independent  (``irreducible")  first  class constraints.  The
purpose of this  paper  is  to  investigate equivalence in the case of
first  class  reducible  constraints,  for  which  some  constraints  are
consequences of the others. The reducible case raises new problems
with respect to the
irreducible one. For instance, in order to get a consistent Dirac
quantization, it is necessary not only that the constraints
remain first-class quantum
mechanically, but also that they remain dependent. Otherwise, the number of
degrees of freedom in the classical and quantum theories would be different.
Furthermore, in the BRST formalism, ghosts of ghost are necessary besides
the standard  ghosts,  and  it is of interest to understand their role in
definite quantum models.

As the general question of equivalence is quite intricate, we restrict in this
paper the analysis to reducible first class systems with constraints that are
linear, homogeneous in the momenta. This case is already of interest
since  it  covers $p$-form gauge fields  and  illustrate  very  well  the
crucial  role  played  by  the  ghosts  of  ghosts.    The  corresponding
irreducible models have been investigated in Ref.$[2,7,9]$.

In the framework of the quantization rules where the
physical wave functions are taken to be densities of
weight one -half in the configuration space, we show that the three
methods of quantization yield the same physical spectrum,
provided one transforms appropriately
the Dirac wave functions under a redefinition of the constraints and of the
reducibility  coefficients.    In  order  to  get  a    consistent  Dirac
quantization, we also find it necessary to
correct the naive Dirac operator condition by an extra  term.  This extra
term,  as  well  as  the  transformation  properties  of  the Dirac  wave
functions,  are quite natural from the BRST point of view.   Our  results
generalize  to  the  reducible case those derived by Tuynman in Ref.$[9]$
for irreducible constraints.

Our paper  is  organized  as  follows.   In the next section, we describe
explicitly the models  considered  in  this  paper.    We then derive the
classical BRST charge that captures all the identities fulfilled by those
models (Section 3).  We  turn  next  to  the  quantization of the models,
first along the lines of the  reduced phase space method (Section 4), and
then along those  of  the Dirac approach (Section 5).  We find it crucial
to improve the naive Dirac quantum constraints by an appropiate term that
makes them anomaly free, and we derive this term by geometric arguments.
Section  6 establishes  the equivalence of the reduced phase space and
Dirac methods by developing further the geometric
interpretation.  The BRST quantization and its equivalence with the other
methods of quantization are given in Section 7.  The  key  role played by
the ghosts and ghosts of  ghosts  is  particularly  stressed  since  they
precisely yield the anomaly cancelling term  of  the  Dirac  quantization
method. Finally, Section 8 is devoted to concluding comments.
\vskip4pc
\centerline{\bf 2. THE MODELS}
\vskip2pc
The systems considered  in this paper are  described by $n$
pairs of canonically conjugate variables $(q^i,p_i)$. They are subject
to $m_0$ bosonic constraints,
$$\G{a}(q^i,p_j)=0,\ \ \ \ \ \  a_0=1,...,m_0,\eqno(2.1)$$
which we take to be linear in the momenta
$$\G{a}(q^i,p_j)=\x{j}{a}(q^i) p_j.\eqno(2.2)$$
The constraints are first class, i.e.
$$\left\{\G{a},\G{b}\right\}=\C{c}{a}{b}\G{c},\eqno(2.3)$$
where $\{,\}$ stands for the Poisson bracket in the phase
space spanned  by the variables $(q^i,p_i)$. Since  the constraints are linear
in the momenta, the structure  functions $\C{c}{a}{b}$ can be taken to
depend only on the
coordinates $q^i$. Furthermore, the gauge  transformation of a function
$f({\bf q})$ defined on the configuration space $\cal Q$ ,is
$$\delta f({\bf q})=\epsilon^{a_0}\{f,\G{a}\}= \epsilon ^{a_0}\x{i}{a}
{\partial f\over\partial q_i}=\epsilon ^{a_0}\vx{a}(f),\eqno(2.4)$$
and depends also only on ${\bf q}$. The vector fields $\vx{a}$ define
the  gauge transformations in the configuration space and are
tangent to the gauge orbits. By inserting eq.(2.2) in eq.(2.3)
we obtain
$$2\ {\bx{j}{b}}{\x{i}{a}}_{],j}=
{{\x{i}{a}}_{,j}}{\x{j}{b}}-{\x{i}{b}}_{,j}{\x{j}{a}}=
{\C{c}{a}{b}}{\x{i}{c}},\eqno(2.5.a)$$
where $_{,j}$  denotes  differentiation  with  respect  to  $q^j$.    The
equation $(2.5.a)$ can be rewritten as
$${{\cal L}_{\vx{a}}}\vx{b} = [\vx{a},\vx{b}] = -\C{c}{a}{b}\vx{c},
\eqno(2.5.b)$$
where $[\ ,\  ]$  is the Lie bracket and ${{\cal L}_{\vx{a}}}$ is the Lie
derivative operator along $\vx{a}$.

The gauge transformations generated by the constraints are said
to be  reducible when  there exist functions
${\Z{a}{a}{0}{1}}\not\approx 0$
such that
$$\Z{a}{a}{0}{1}\G{a}=0.\eqno(2.6)$$
Because the constraints are linear and homogeneous in the momenta, we may
assume  the reducibility functions $\Z{a}{a}{0}{1}$  to  depend  only  on
$q^i$; eq.$(2.6)$ is then equivalent to
$$\Z{a}{a}{0}{1}({\bf q})\ \vx{a}({\bf q}) = 0,\ \ \ \ \ \ \
a_1=1,...,m_1.\eqno(2.7)$$
The functions $\Z{a}{a}{0}{1}$  are required to exhaust all
the relations among the fields $\vx{a}$.

It might happen that
the set $\{\Z{a}{a}{0}{1}\}$ is overcomplete, i.e., there exists
a set of functions $\Z{a}{a}{1}{2}$ such that
$$\Z{a}{a}{1}{2}\Z{a}{a}{0}{1}\approx 0. \eqno(2.8)$$
Eq. $(2.8)$ means that $\Z{a}{a}{1}{2}\Z{a}{a}{0}{1}$ can be
written as a combination  of the constraints. Again, the functions
$\Z{a}{a}{1}{2}$ may be taken to depend only on ${\bf q}$.
Since the constraints depend
on the momenta but the $Z$'s do not, the only possibility is that eq.$(2.8)$
is valid strongly. In general one finds a tower  of reducibility equations:
$$\Z{a}{a}{k}{{k+1}}\Z{a}{a}{{k-1}}{k}=0,\ \ \ \
a_k=1,...,m_k,\ \ \ \ k=1,...,L-1.\eqno(2.9.a)$$
The tower stops with functions $\Z{a}{a}{{L-1}}{L}({\bf q})$ that are
linearly independent,
$$\lambda^{a_L}({\bf q})\ \Z{a}{a}{{L-1}}{L}=0  \Rightarrow
\lambda^{a_L} = 0. \eqno(2.9.b)$$
The theory has then order of reducibility equal
to $L$. The number of independent gauge generators is:
$$m=\sum^{L}_{k=0}(-)^{k}m_k. \eqno(2.10)$$
It will be convenient to choose
the $Z$'s such that $${\Z{a}{a}{{k-1}}{k}}^* = (-)^k
\Z{a}{a}{{k-1}}{k}.\eqno(2.11)$$
\bigskip
Non linearly independent gauge generators appear in a
physical theory when one cannot isolate a subset of independent
constraints without violating explicit covariance, locality, or global
conditions. A well known example is the case of a $p$-form gauge field
${\bf A}=(1/p\ !)\ A_{{i_1...i_p}}dx^{i_1}\wedge...\wedge d{x}^{i_p}$,
the canonically conjugated pairs
being $\left(A_{i_1...i_p}(x),\pi^{i_1...i_p}(x)\right)$.
For such a field  the constraints are:
$${\pi^{{i_1...i_p}}}_{,i_{1}}=0,$$
(the generalization of the Gauss law) which are not
independent because of the antisymmetrization in the
indices $i_1,...,i_p$
(${\pi^{i_1 i_2 ... i_p}}_{,i_1 i_2}\equiv 0$).
\vskip4pc
\centerline{\bf 3. THE CLASSICAL BRST GENERATOR}
\vskip2pc
{\bf 3.1 IDENTITIES}
\vskip1pc
The  functions  $\x{i}{a} ({\bf q})$ and $\Z{a}{a}{{k-1}}{k}  ({\bf  q})$
fulfill a series of identities that can be derived from eqs.$(2.5)$ and
$(2.6)$ by differentiation and use of the symmetry of  the second partial
derivatives. For instance, the Jacobi identity,
$$\{\{G_{[a_0},G_{b_0}\},G_{{c_0]}}\}=0,$$
leads to the equation:
$$\left(\bx{i}{a}{\C{e}{b}{c}}_{],i}-
\C{d}{[a}{b}\C{e}{c}{]d}\right)\G{e}=0.$$
For a reducible theory, this means that there exist
functions ${\M{a}{a}{b}{c}{1}{0}}({\bf q})$ such that
$$\bx{i}{a}{\C{e}{b}{c}}_{],i}=\C{d}{[a}{b}\C{e}{c}{]d}
- {2\over 3} \M{a}{a}{b}{c}{1}{0} \Z{e}{a}{0}{1},
\eqno(3.1)$$
the last term being not present when the theory is irreducible.
\medskip
Similarly, by differentiating the identity $(2.7)$ along the
orbits, one gets
$${\x{i}{b}{\left(\Z{a}{a}{0}{1}\x{j}{a}\right)},_i}=0,$$
or, in terms of the vector fields $\vx{a}$,
$$\left[{\cal L}_{\vx{b}}\left(\Z{a}{a}{0}{1} \vx{a}\right)\right]^j
= \x{i}{b}\left(\Z{a}{a}{0}{1} \x{j}{a}\right)_{,i}-
{\x{j}{b}}_{,i} \Z{a}{a}{0}{1} \x{i}{a}=0,$$
i.e.
$$\x{i}{b} {\Z{a}{a}{0}{1}}_{,i}\ \x{j}{a}+
\Z{a}{a}{0}{1} \left(\x{i}{b} {\x{j}{a}}_{,i}-
{\x{j}{b}}_{,i} \x{i}{a}\right)=0.$$
Because of eq.(2.5.a), this is equivalent to
$$\left(\x{i}{b} {\Z{a}{a}{0}{1}}_{,i} + \Z{c}{a}{0}{1} \C{a}{c}{b}
\right) \x{j}{a}=0.$$
The completeness of the functions $\Z{a}{a}{0}{1}$ implies then the existence
of

functions $\D{b}{b}{a}{1} ({\bf q})$  such that:
$$\x{i}{b} {\Z{a}{a}{0}{1}}_{,i} + \Z{c}{a}{0}{1} \C{a}{c}{b} =
\D{b}{b}{a}{1} \Z{a}{b}{0}{1}.\eqno(3.2)$$
\bigskip
If one contracts this identity with $\Z{b}{c}{0}{1}$, sums over $b_0$ and
uses eq.$(2.7)$, one gets
$$\Z{b}{c}{0}{1} \Z{c}{a}{0}{1} \C{a}{c}{b} =
\Z{b}{c}{0}{1} \Z{a}{b}{0}{1} \D{b}{b}{a}{1}$$
The symmetric part of this equation in $a_1,c_1$, reads
$$\D{b}{b}{{(a}}{1} \Z{b}{c}{0}{{1)}}
\Z{a}{b}{0}{1} = 0.$$
The completeness  of  $\Z{a}{b}{0}{1}$  implies  then  the  existence  of
functions $\B{b}{d}{b}{2}{1} ({\bf q})$ such that
$$\D{a}{b}{c}{1} \Z{b}{d}{0}{1} + \D{a}{d}{b}{1}
\Z{b}{c}{0}{1} = \B{b}{d}{c}{2}{1} \Z{a}{b}{1}{2},\eqno(3.3)$$
the right-hand side  of this equation being absent for reducible theories
of order $L=1$.
\vskip3pc
{\bf 3.2. THE BRST GENERATOR}
\vskip1pc
The identities $(3.1)$-$(3.3)$ are  only  a  few  of  a long list, which
can be obtained by further differentiation.
The  most  powerful way to capture
all these identities is to introduce  the  BRST  generator  $^{15}$.  The
identities are then contained in a unique equation, namely
$$\left\{\Omega ,\Omega\right\} = 0,\eqno(3.4)$$
where $ \Omega =\Omega ( q^i,p_j,\e{a}{k},\P{b}{k})$,
the BRST generator, is a fermionic function in an extended
phase space ${\cal E}$ including canonically conjugate pairs of
ghosts $(\e{a}{k},\P{a}{k})$, $k=0,...,L$, besides the original canonical
variables,
$$\left\{\P{a}{k},\e{b}{k}\right\} = - \delta ^{b_k}_{a_k},\eqno(3.5.a)$$
$$gh(\e{a}{k}) = - gh(\P{a}{k}) = k+1.\eqno(3.5.b)$$

In eq.$(3.4)$  the bracket is the Poisson bracket in ${\cal E}$; we remark
that the nilpotency condition $(3.4)$  is not trivial because
the Poisson bracket for fermionic quantities is symmetric.

The BRST generator is unique, up to canonical transformations in ${\cal E}$.
For reducible theories, $\Omega$ has the form$^{16,17}$
$$\Omega = \e{a}{0} \G{a} + \sum^{L-1}_{k=0}\e{a}{{k+1}} \Z{a}{a}{k}{{k+1}}
\P{a}{k} + ``more",\eqno(3.6)$$
where $``more"$ does not contains terms of the already indicated form.

In our case the $G$'s  and the  $Z$'s are bosonic; in order
that  $\Omega$ be fermionic, the ghosts belonging to an even  generation
$\left(\e{a}{0},\P{a}{0},\e{a}{2},\P{a}{2},...\right) $ must be fermionic,
while
those belonging to an  odd generation must be bosonic.
Due to the choice $(2.11)$, $\Omega$  turn out to be real if
$${\e{a}{k}}^* = \e{a}{k},\ \ \ \ \ \ \ \ {\P{a}{k}}^* = (-)^{k+1} \P{a}{k}.
\eqno(3.7)$$
\smallskip
The generator $\Omega $ can be built by means of a recursive method
(see Ref.[1]):
$$ \Omega = \sum_{p\geq 0} \Omega\!\!\!\!\!^{^{^{(p)}}} , \eqno (3.8)$$
where  $ \Omega\!\!\!\!\!\!\! ^{^{^{(p+1)}}} $  solves the equation
$$ \delta\ \ \Omega\!\!\!\!\!\!\!^{^{^{(p+1)}}} +
\Delta\!\!\!\!\!^{^{^{(p)}}} =0,\eqno (3.9)$$
with
$$ \Delta\!\!\!\!\!^{^{^{(p)}}} = {1\over
2}\sum^p_{k=0}\left\{\ \Omega\!\!\!\!\!\!\!^{^{^{(p-k)}}},
\Omega\!\!\!\!\!^{^{^{(k)}}}\right\}_{orig} + {1\over
2}\sum^p_{k=1}\sum^{k-1}_{s=0}\left\{\ \ \ \
\Omega\!\!\!\!\!\!\!\!\!\!\!\! ^{^{^{(p-k+s+1)}}},
\Omega\!\!\!\!\!^{^{^{(k)}}}\right\}_{\e{a}{s},\P{a}{s}},\eqno(3.10.a)$$
$$ \Omega\!\!\!\!\!^{^{^{(0)}}}
={\e{a}{0}}\G{a}.\eqno(3.10.b)$$
In  eq.$(3.10)$ $\{\  ,\  \}_{orig}$  is the bracket with  respect  to  the
original  canonical  variables,  while $\{\ ,\ \}_{\e{a}{s},\P{a}{s}}$ is
the  bracket   with  respect  to  the  pair  $(\e{a}{s},\P{a}{s})$.    In
eq.$(3.9)$ $ \delta $ is the Koszul-Tate operator. In
our case $ \delta $ reads explicitly
$$ {\delta q^i}=0, \ \ \ \ \ \ \ {\delta p_j}=0 , \eqno(3.11.a)$$
$$ \delta {\e{a}{0}} =0,\ \ \ \ \ \ \ \delta \P{a}{0}=-\G{a}\eqno(3.11.b)$$
$$\delta \e{a}{k}=0\ \ \ \ \ \ \delta
\P{a}{k}=-\Z{a}{a}{{k-1}}{k}\P{a}{{k-1}}, \ \ \ \ \ k=1,...,L,\eqno(3.11.c)$$
and is clearly nilpotent ($\delta ^2=0$), because the reducibility
eq.$(2.9.a)$ holds strongly when
the constraints are linear in the momenta (in the general case, additional
terms are needed in $(3.11)$ to achieve nilpotency).
\vskip1pc
The existence of  $\Omega$ is established in Ref.[17].  Its explicit form
for arbitrary $L$ is  cumbersome  and  will not be needed here.  We shall
only need:  $(i)$ the crucial fact that $\Omega$ is linear in the momenta
$(p_i, \P{a}{k})$ (Proposition 1);  and $(ii)$ the identities in
Propositions 2, 3 and 4 below.
\vskip3pc
{\bf Proposition 1.} In the case
of constraints linear in the momenta $p_j$, the BRST generator $\Omega$ can be
taken to be linear in the momenta $ (p_j,\P{a}{k})$.
\smallskip
{\bf Proof.} One has
$$\Delta\!\!\!\!\!^{^{^{(0)}}}={1\over2}\left\{{\Omega\!\!\!\!\!^{^{^{(0)}}},
\Omega\!\!\!\!\!^{^{^{(0)}}}}\right\}_{orig}=
{1\over2} \e{a}{0} \e{b}{0} \left\{{\G{a},\G{b}}\right\}=
{1\over2} \e{a}{0} \e{b}{0} \C{c}{a}{b} \G{c} $$
$$\Rightarrow \Omega\!\!\!\!\!^{^{^{(1)}}}={1\over2}\e{a}{0}\e{b}{0}
\C{c}{a}{b}\P{c}{0}.$$
So let us suppose that all the $ \Omega\!\!\!\!\!^{^{^{(k)}}} $'s
for $ k \leq  p $  are  linear in the momenta (this is true for $  p=1 $ ).
Then $\Delta\!\!\!\!\!^{^{^{(p)}}} $  is linear in the momenta from $(3.10)$.
Because of the definition $(3.11)$ for $\delta$,
we find then that $ \Omega\!\!\!\!\!\!\!^{^{^{({p+1})}}} $  in eq.$(3.9)$
can be taken to be linear in the momenta.$\diamondsuit$

\bigskip
By expanding out  $\{\Omega,\Omega\}=0$,  one
finds at the lowest orders in the ghosts and their momenta the identities
$(2.5)$,  $(2.7)$  and    $(2.9)$, since  $\C{c}{a}{b}  /2$  is  the
coefficient of $\e{a}{0} \e{b}{0} \P{a}{0}$ in $\Omega$, as we just got.
Similarly, by calling $\M{a}{a}{b}{c}{1}{0} /3$,  the coefficient of
$\e{a}{0} \e{b}{0} \e{c}{0} \P{a}{1}$,
one finds the identity $(3.1)$. In addition one has:
\bigskip
{\bf Proposition 2.} There exist functions $\D{c}{b}{a}{k}$ such that
$$\x{i}{b} {\Z{a}{a}{{k-1}}{k}}_{,i}+ \D{a}{b}{c}{{k-1}}
\Z{c}{a}{{k-1}}{k} =
\D{c}{b}{a}{k} \Z{a}{c}{{k-1}}{k},\ \ \ \ \ \ k=1,...,L.\eqno(3.12)$$

\vskip2pc
{\bf Proposition 3.} There exist functions $\B{b}{c}{a}{{k+1}}{k}$ such that
$$\D{a}{b}{c}{k} \Z{b}{d}{0}{1} - \B{a}{d}{b}{k}{{k-1}}
\Z{b}{c}{{k-1}}{k} = \B{b}{d}{c}{{k+1}}{k} \Z{a}{b}{k}{{k+1}}
\ \ \ \ \  k=1,...,L-1,\eqno(3.13.a)$$

$$\D{a}{b}{c}{L} \Z{b}{d}{0}{1} =
\B{a}{d}{b}{L}{{L-1}} \Z{b}{c}{{L-1}}{L}.\eqno(3.13.b)$$
\vskip2pc
{\bf Proposition 4.} There exist functions $\M{d}{a}{b}{c}{k}{{k-1}}$ such
that
$$\eqalign{{\bx{i}{a} \D{c}{b}{]a}{k}}_{,i}=-{1\over 2}\C{e}{a}{b}
\D{c}{e}{a}{k}&+{\D{d}{[a}{{_{^|}a}}{k}}_{^|} \D{c}{b}{{]d}}{k} +
\M{c}{a}{b}{f}{k}{{k-1}} \Z{f}{a}{{k-1}}{k}\cr &+
\M{f}{a}{b}{a}{{k+1}}{k} \Z{c}{f}{k}{{k+1}},
\ \ \ \ \ \ \ \ \ k=1,...,L-1.\cr}\eqno(3.14.a)$$
\medskip
$$\bx{i}{a} {\D{c}{b}{{]a}}{L}}_{,i} = -{1\over 2} \C{e}{a}{b}
\D{c}{e}{a}{L} + {\D{d}{{[a}}{{_{^|}a}}{L}}_{^|} \D{c}{b}{{]d}}{L} +
\M{c}{a}{b}{f}{L}{{L-1}} \Z{f}{a}{{L-1}}{L}.\eqno (3.14.b) $$
\vskip2pc
The    proof  of  these  propositions  goes  as    follows.        Define
$\D{c}{a}{b}{0}$ $\equiv$ $-\C{c}{a}{b}$, and $\B{b}{d}{c}{2}{1}/2$ to be
the coefficient of $\e{d}{1} \e{c}{1} \P{b}{2}$ in $\Omega$.
Similarly, define $-\D{b}{b}{a}{k}$,
$\B{b}{d}{c}{{k+2}}{{k+1}}$
and $\M{a}{a}{b}{c}{{k+1}}{k}$ ($k\ne 0$) to be the
respective coefficients of $\e{b}{0}  \e{a}{k}  \P{b}{k}$,
$\e{d}{1} \e{c}{{k+1}} \P{b}{{k+2}}$ and
$\e{a}{0} \e{b}{0} \e{c}{k} \P{a}{{k+1}}$ in $\Omega$.
Then, the vanishing of the
coefficients  of $\e{a}{k}  \e{b}{0}  \P{a}{{k-1}}$,  $\e{d}{1}  \e{c}{k}
\P{a}{k}$,    and    $\e{a}{k}    \e{b}{0}    \e{a}{0}    \P{c}{k}$    in
$\{\Omega,\Omega\}$ yield respectively $(3.12)$, $(3.13)$, and $(3.14)$.
Note  that since the ghosts  $\e{a}{1}$  are  bosonic,  the  coefficients
$\B{a}{c}{d}{2}{1}$ are symmetric in $(c_1,d_1)$,
$$\B{a}{c}{d}{2}{1} = \B{a}{d}{c}{2}{1}.\eqno(3.15)$$
Note  also that the identity $(3.12)$ reduces  to  the  identity  $(3.2)$
for  $k=1$, and $(3.13)$ (with $\B{c}{a}{b}{1}{0}=-\D{c}{b}{a}{1}$)
becomes $(3.3)$ for $k=1$.
\bigskip
It is of  interest  to  write  explicitly the BRST charge for a reducible
theory of order $L=1$. One gets
$$\Omega = \e{a}{0} \G{a} + \e{a}{1} \Z{a}{a}{0}{1}
\P{a}{0}  +  {1\over 2}\  \e{a}{0}\  \e{b}{0}\  \C{c}{a}{b}\  \P{c}{0}  -
\e{b}{0}\ \D{a}{b}{b}{1}\  \e{b}{1}\  \P{a}{1}  +  {1\over  3}  \e{a}{0}\
\e{b}{0}\ \e{c}{0}\ \M{a}{a}{b}{c}{1}{0}\ \P{a}{1}.\eqno(3.16)$$
\vskip4pc
\centerline{\bf 4. REDUCED PHASE SPACE QUANTIZATION}
\vskip2pc
Because the gauge transformations are defined within the space ${\cal Q}$ of
the  $q$'s,  one  can  introduce  the  reduced  configuration  space  ${\cal
Y \equiv Q/G}$ as the quotient of the configuration space ${\cal Q}$ by the
gauge  orbits  in  ${\cal  Q}$.     Let  $y^\alpha$, $\alpha  =1,...,N$,  be
coordinates in the reduced configuration space.   $N$ is equal to $n$ minus
the number $m$ of independent constraints. One has
$$\{y^\alpha({\bf q}), \G{a}\} = 0.\eqno(4.1)$$
Let  $\pi_\alpha({\bf q,p})$ be the gauge invariant  momenta  conjugate  to
$y^\alpha$,
$$\{\pi_\alpha, \G{a}\} \approx 0,\ \ \ \ \  \  \ \ \{y^\alpha, \pi_\beta\}
\approx \delta^\alpha_\beta.\eqno(4.2)$$
The  variables  $y^\alpha$ and $\pi_\alpha$ define a standard unconstrained
system,  the  ``reduced  system" associated with the original gauge system.
The  reduced  phase  space quantization consists in quantizing this reduced
system without worring about its origins.

So, let us consider a non-constrained system described
classically by coordinates and momenta $(y^\alpha ,\pi
_\alpha )$, $\alpha =1,...,N$. At a given time, the quantum state of
the system is given by a wave function $\varphi (y^\alpha
)$ belonging to a Hilbert space. It is convenient to define the inner
product  in this space as:
$$(\varphi,\psi) = \int d^Ny\ \ \varphi^*(y^\alpha )\
\psi(y^\alpha ),\eqno(4.3)$$
In  order that the inner product $(4.3)$ be invariant
under  coordinate changes, the wave functions must behave as
scalar densities of weight $1/2$:

$$y^\alpha \longrightarrow y'^\alpha = y'^{\alpha}(y^\alpha )
\Rightarrow \varphi(y^\alpha )\longrightarrow\varphi '(y'^{\alpha}) =
\left\vert  {\partial y^\alpha \over \partial y'^{\alpha}}\right\vert^{1/2}
\varphi \left(y^\alpha (y'^{\alpha})\right).\eqno(4.4)$$
The product $\varphi^* \psi$ is then a density of weight 1, i.e.,
defines a
$N$-form  in  ${\cal   Y}$.    Since  the  integral  of  a  $N$-form  over  a
$N$-dimensional manifold is intrinsically  defined  (does  not  require  an
extra integration measure), the convention  of taking the wave functions to
be densities of weight 1/2 is  convenient  in  the  absence  of  a  natural
integration measure \footnote{{$^\dagger$}}{In practice, however, the manifold
comes  equipped  with  an
integration measure $\nu$.  For instance, it is a Riemannian manifold and
$\nu=g^{1/2}$.  In  that  case, one can replace the wave functions by
scalars, by redefining them as  $\varphi\rightarrow\nu^{-1/2} \varphi$.  Of
course, this procedure also requires a  redefinition  of  the  operators in
order to keep the matrix elements unchanged.}.

The observables that are linear in the  momenta  $\pi_\alpha$  conjugate to
$y^\alpha$,
$${\bf a} = a^\alpha({\bf y})\ \pi_\alpha\ \ \  \  \  \  \  \  \ \ \ {\rm
(classically),}\eqno(4.5)$$
possess a natural geometric interpretation since they define  vector fields
on the manifold of the $y$'s. Their quantum version reads
$${\bf a} =   {1\over  2}  \left(a^\alpha({\bf  y})\  \pi_\alpha  +
\pi_\alpha\
a^\alpha({\bf y})\right),\eqno(4.6)$$
and  is  formally   Hermitian  for  the  scalar  product  $(4.3)$  whenever
$a^\alpha$ is real.  The action of ${\bf a}$ on a wave function yields $-i$
times its Lie derivative (as a density of weight 1/2),
$$({\bf a}\ \varphi) ({\bf y}) = -i {\cal L}_{\bf a}\
\varphi\ \ \ \ \ \ \ \ \ \ \ \ \ \ \ \eqno(4.7.a)$$
$$\ \ \ \ \ \ \ \ \ \ \ \ \ \ \ \ \ \ \ =  -i \left(a^\alpha\
\varphi_{,\alpha} + {1\over  2}  a^\alpha_{,\alpha}\
\varphi\right).\eqno(4.7.b)$$
\vskip4pc
\centerline{\bf 5. DIRAC QUANTIZATION}
\vskip2pc
We now turn to the Dirac quantization, where the wave funtions are taken to
depend on all the coordinates $q^i$ and not  just  on  the  gauge invariant
ones.  In order to remove the unphysical degrees of freedom, one imposes on
the physical states the condition
$$\hG{a}\ \psi({\bf q}) = 0,\eqno(5.1)$$
where $\hG{a}$ is the realization of each constraint as an
operator in the space of the wave functions. Since the quantum
realization of any classical function is ambiguous (factor ordering problem),
we should carefully define the operator $\hG{a}$.
Because the constraints are linear in the momenta, it is natural to take
$\hG{a} \psi$ to be the Lie derivative of $\psi$ along  $\vx{a}$.  However,
the Lie derivative of $\psi$ is ill-defined as long as one does not give the
transformation  rules  for  $\psi$.    So the question is:  which object is
$\psi$?  In order to gain insight into this question, let us first consider
the simple abelian case.
\vskip2pc
\vfill
\eject
{\bf 5.1. ABELIAN CASE}
\vskip1pc
Let  us  thus    assume   that  the  coordinates  $q^i$  can  be  split  as
$q^i\equiv(y^\alpha,Q^A)$, in such a way that the constraints $\G{a}\approx
0$ are equivalent to
$$G_A\equiv P_A \approx 0,\eqno(5.2)$$
$P_A$ being the momenta conjugate to $Q^A$.
Locally, this can always be  achieved  $^{1}$.  The reduced phase space for
the system is then the space  of  the $y^\alpha$ and the $\pi_\alpha$.  The
discussion of Section 4 shows that the  physical wave functions depend only
on $y^\alpha$, i.e.  are annihilated by ${\hat  P}_A$.    Furthermore,  the
scalar product $(4.3)$ can be rewritten as
$$(\varphi, \psi) = \int d^Ny\ d^mQ \prod^m_{A=1}\delta(Q^A)\ \
\varphi^*(y^\alpha)\ \psi(y^\alpha ).\eqno(5.3)$$
This expression is invariant if we transform the wave functions not only as
densities of weight 1/2 under changes of the physical coordinates $y^\alpha$
(as we already pointed out), but also as {\it scalars  under changes of the
pure gauge coordinates} $Q^A$.

This asymmetric behaviour of the wave functions under change of coordinates
is  undesirable  since  in  practice,  one cannot split the $q^i$ into  the
$y^\alpha$  and  the  $Q^A$.    It  is  thus  necessary  to reformulate the
transformation properties of the wave functions in a manner that treats the
coordinates more uniformly. To that end, we rewrite (5.3) as
$$(\varphi, \psi) = \int  d^N y\ d^m Q\ \prod^m_{A=1}
\delta\left(\chi^A\right)\ \left\vert det\{\chi^A, G_B\}\right\vert\
\varphi^*(y^\alpha)\ \psi(y^\alpha ),\eqno(5.4)$$
where $\chi^A=0$ define  good  gauge  conditions.    The  inclusion  of the
``Faddeev-Popov" determinant $det\{\chi^A, G_B\}$  makes  (5.4) independent
of $\chi^A$, and equal to  $(5.3)$  (to  see  this, take $\chi^A=Q^A$).  We
then observe that under a redefinition of the $Q$'s,
$$Q^A \longrightarrow Q'^A = Q'^A (Q^A,y^\alpha),\eqno(5.5)$$
the momenta $P_A$ conjugate to $Q^A$ transform as
$$P_A = {\partial Q'^B\over\partial Q^A} \ P'_B.\eqno(5.6)$$
Hence, the original constraints $(5.2)$ are not identical with $P'_A$ but
differ  from  the  constraints  $G'_A\equiv P'_A\approx 0$ adapted to the
$Q'^A$-description by a $q^i$-dependent linear transformation.  In others
words, in  order  to  reach the description of the system in terms of
the new pure  gauge  variables  $Q'^A$,  one  must  supplement the change
$(5.5)$ by a redefinition of the constraints.

Now, the scalar product $(5.4)$ is invariant under
$$G_A \longrightarrow G'_A = A_A\ ^B\ G_B,\eqno(5.7)$$
if and only if  the wave functions transforms as densities of weight $-1/2$
for $(5.7)$.  Accordingly, we  shall  postulate  that  the  wave  functions
transforms as densities of weight 1/2 for
$$q^i \longrightarrow  q'^i = q'^i (q^j)\eqno(5.8)$$
and as  densities of  weight $-1/2$ for $(5.7)$,
$$\psi({\bf q}) \longrightarrow \psi'({\bf q}') = \mid det
A\mid ^{-1/2} \left\vert {\partial{\bf q}\over
\partial{\bf q}'}\right\vert^{1/2}\ \psi({\bf q}).\eqno(5.9)$$
This automatically guarantees that $\psi$ is  a  scalar  under  changes  of
coordinates  along  the  gauges  orbits,  since  the  redefinition  of  the
constraints  (5.6)  ($P_A \rightarrow P'_B=(\partial Q^A/\partial Q'^B)
P_A$) compensates the Jacobian coming from the density weight of $\psi$,
$$\psi({\bf  y,Q})   \longrightarrow  \psi'({\bf  y,Q}')  =\left\vert  det
{\partial {\bf Q}\over\partial {\bf Q}'}\right\vert ^{-1/2}\  \left\vert  det
{\partial {\bf Q}\over\partial {\bf Q}'}\right\vert ^{1/2} \psi({\bf y,Q}).$$
The conclusion is  that  in  order  to treat uniformly the coordinates, one
must require the Dirac  wave  functions  to  transform  non trivially as in
$(5.9)$ under a redefinition of the constraints.
\vskip2pc
{\bf 5.2. GENERAL CASE - DEFINITION OF LIE DERIVATIVE OF $\psi$}
\vskip1pc
In the general case  $(2.1)$-$(2.9)$  of  reducible constraints, not only
can one ``rotate" the constraints,
$$\G{a}\longrightarrow    G'_{a_0}({\bf q})    = \A{a}{b}{0}({\bf    q})\
\G{b}({\bf q}),\eqno(5.10)$$
but    one    can   also    transform    the    reducibility    functions
$\Z{a}{a}{{k-1}}{k}$ as
$$\Z{a}{a}{{k-1}}{k}({\bf q}) \longrightarrow Z'^{a_{k-1}}_{a_k}({\bf q}) =
\A{a}{b}{k}({\bf q})\ \Z{b}{b}{{k-1}}{k}({\bf q})
\ \Ai{b}{a}{{k-1}}({\bf q}).\eqno(5.11)$$
We shall generalize the previous  transformation  laws  by requiring that
the Dirac wave functions transform as  \footnote{{$^\dagger$}}{Given the new
and the old constraints and reducibility functions,  one  cannot read off
uniquely  the  matrices  $\A{a}{b}{k}$.   For instance, $\A{a}{b}{0}$  is
determined up to $\mu_{_0}^{b_1} \Z{b}{b}{0}{1}$.  However, it is easy to
convince oneself that this ambiguity in the $A$'s does not  affect $\psi'$
in $(5.12)$.}
$$\psi({\bf q}) \longrightarrow \psi'({\bf q}')
=\prod^L_{k=0} \left\vert det \A{a}{b}{k}\right\vert ^{(-)^{k+1}/2}
\ \ \ \left\vert {\partial{\bf q}\over \partial{\bf
q}'}\right\vert ^{1/2} \psi({\bf q}).\eqno(5.12)$$
under $(5.10)$, $(5.11)$ and coordinate transformations $(5.8)$.  The law
$(5.12)$ reduces to $(5.9)$ for irreducible constraints.

We shall show in Section 6 that this is the correct choice in that it
yields a Dirac quantization  equivalent  to  the reduced phase space one.
In this section, we shall  verify  that  the  Dirac quantization based on
$(5.12)$ is consistent.  Namely, that  it  leads  to  quantum constraints
$(5.1)$ that are still first class,
$$\left[\hG{a},\hG{b}\right] = i {\hat C}^{c_0}_{a_0 b_0} \hG{c},\eqno(5.13)$$
(in that order) and that fulfill
$${\hat Z}^{a_0}_{a_1}\ \hG{a} = 0,\eqno(5.14)$$
(in  that  order).    The  equation $(5.13)$  expresses  the  absence  of
anomalies  and  guarantees  the  compatibility of the quantum  conditions
$\hG{a}  \psi=0$.    The  equation  $(5.14)$  means  that  among  $\hG{a}
\psi=0$, there are only $m$ independent equations.

To  prove  $(5.13)$  and  $(5.14)$,  one must compute the Lie  derivative
${\cal L}_{\vx{a}} \psi$ of the Dirac wave functions.  Now, an infinitesimal
diffeomorphism  generated  by    $\vx{a}$    induces  not  only  a  linear
transformation of the coordinate tangent  frames,
but also a redefinition $(5.10)$,
$(5.11)$ of the constraints and  of  the  reducibility functions.  Indeed
one gets from $(2.5)$ and $(3.12)$
$${{\cal L}_{\vx{a}}}\G{b} = -\C{c}{a}{b}\G{c},\eqno(5.15.a)$$
$$\x{i}{a}\  {\Z{a}{b}{{k-1}}{k}}_{,i} = \D{c}{a}{b}{k} \Z{a}{c}{{k-1}}{k}
-\D{a}{a}{c}{{k-1}}\ \Z{c}{b}{{k-1}}{k}.\eqno(5.15.b)$$
Therefore  the  Lie  derivative of $\psi$  involves  not  only  the  term
$(1/2){\x{i}{a}}_{,i}  \psi$,  reflecting  the  weight  of $\psi$  under
coordinate  changes,  but  also  terms  arising from its  variance  under
$(5.10)$ and $(5.11)$.  Let $R$ be the point of coordinates $q^i$ and $S$
the point of coordinates
$$q^i + \epsilon\ \x{i}{a}.\eqno(5.16)$$
(for fixed $a_0$). The Jacobian matrix of $(5.16)$ is
$$\delta^i_j - \epsilon\ {\x{i}{a}}_{,j},$$
while the matrices $\A{a}{b}{k}$ induced by $(5.16)$ are
$$\A{b}{c}{0} = \delta^{c_0}_{b_0} - \epsilon\ \C{c}{a}{b},$$
$$\A{b}{c}{k} = \delta^{c_k}_{b_k} + \epsilon\ \D{c}{a}{b}{k},\ \ \ \ \ \
\ \ k=1,...,L,$$
($G'_{b_0}=\G{b}+\epsilon\ {\cal L}_{\vx{a}} \G{b}$,
$\ \ \ \ Z'^{c_{k-1}}_{b_k}=\Z{c}{b}{{k-1}}{k}+\epsilon\ \x{i}{a}\
{\Z{c}{b}{{k-1}}{k}}_{,i}$).
\bigskip
Thus, one gets
$$\psi_S = \psi_R + \epsilon\ \x{i}{a}\ \psi_{,i},$$
$$\psi_{R\rightarrow   S}  =  \left[    1+    {\epsilon\over    2}
\left(-{\x{i}{a}}_{,i}    +    \C{b}{a}{b}    -  \sum^L_{k=1}    (-)^k
\D{b}{a}{b}{k}\right)\right]\ \psi,$$
where $\psi_S$ is the wave function at $S$, $\psi_R$ the wave function at
$R$ and $\psi_{R\rightarrow S}$ the transformed  (at  $S$)  of  the  wave
function at $R$ under the diffeomorphism $(5.16)$  mapping  $R$  in  $S$.
Thus yields finally
$${\cal  L}_{\vx{a}}\  \psi = {\displaystyle\lim_{\epsilon\to 0}}\
{\psi_S - \psi_{R\rightarrow S}\over \epsilon} = \x{i}{a} \ \psi_{,i} +
{1\over 2} \left({\x{i}{a}}_{,i} - \C{b}{a}{b} + \sum^L_{k=1}    (-)^k
\D{b}{a}{b}{k}\right)\ \psi.\eqno(5.17)$$
The  functions  $\C{c}{a}{b}$  and  $\D{c}{a}{b}{k}$  are not  completely
defined by $(2.3)$, $(3.2)$ and $(3.12)$.  However, the alternating trace
in  $(5.17)$  is  unambiguous,  so that $(5.17)$ is well  defined.    (For
instance,

if  $\ \ \ \C{c}{a}{b} \longrightarrow {\bar C}^{c_0}_{a_0 b_0} =
\C{c}{a}{b} +
\mu^{c_1}_{a_0 b_0} \Z{c}{c}{0}{1}, $
\medskip
then  $\D{c}{a}{b}{1} \longrightarrow {\bar  D}^{c_1}_{a_0  b_1}  =
\D{c}{a}{b}{1} - \mu^{c_1}_{a_0 b_0} \Z{b}{b}{0}{1} $, and
\medskip
$\C{c}{a}{b} + \D{c}{a}{b}{1} \longrightarrow {\bar C}^{c_0}_{a_0 b_0} +
{\bar  D}^{c_1}_{a_0  b_1} = \C{c}{a}{b} + \D{c}{a}{b}{1}$).
\vskip2pc
\vfill
\eject
{\bf 5.3. GENERAL CASE - CONSISTENCY OF DIRAC QUANTIZATION}
\vskip1pc
In view of the above discussion, we take the operator $\hG{a}$ in $(5.1)$
to be $-i{\cal L}_{\vx{a}}$, i.e.,
$$\hG{a}\ \psi \equiv -i\ {\cal L}_{\vx{a}}\ \psi = 0,\eqno(5.18)$$
or
$$i  \hG{a}    =    i\    {\hat\xi}^i_{a_0}\    {\hat  p}_i  +  {1\over  2}
\left({{\hat\xi}^i_{a_0 ,i}} - {\hat  C}^{b_0}_{a_0  b_0}  + \sum^L_{k=1}
(-)^k\ {\hat D}^{b_k}_{a_0 b_k} \right).\eqno(5.19)$$
Because the second term in the right-hand side of $(5.19)$ is multiplied by
$\hbar$ (set equal to one  here),  one  can  view  it  as  arising  from an
ordering ambiguity \footnote{{$^\dagger$}}{For any operator $\hat  A$, one has
$\hbar{\hat  A}=\hbar{\hat  A}{\hat  1}=-i{\hat  A} ({\hat q}{\hat  p}-{\hat
p}{\hat q})$.  So, one can always view  $\hbar{\hat  A}$ as arising from an
ordering ambiguity in $-iA(pq-qp)$ (classically equal to zero).}.
In the limit $\hbar\rightarrow 0$, $\hG{a}$  goes  over  into
$\x{i}{a}  p_i$  and  so,  possess the  correct classical limit.

To  verify  the  consistency  of  the  Dirac  quantization,  one must check
$(5.13)$  and $(5.14)$.  This is direct because the Lie derivative $(5.17)$
has the following crucial properties:
\medskip
\line{(i) ${\cal  L}_{\vx{a}}\  \psi$  transforms  as  $\psi$.\hfill (5.20)}

\line{(ii) ${\cal L}_{\mu  \vx{a}}\  \psi  =  \mu\ {\cal L}_{\vx{a}}\ \psi$
for any $\mu({\bf q})$.\hfill (5.21)}

\line{(iii) $[{\cal  L}_{\vx{a}},{\cal  L}_{\vx{b}}]\ \psi =
{\cal  L}_{[\vx{a},\vx{b}]}\  \psi.$\hfill (5.22)}
\medskip
The property $(5.20)$ follows  from  the  definition  $(5.17)$  of  the Lie
derivative since in ${\cal L}_{\vx{a}}\  \psi$,  one  takes the difference of
two objects  that  transform  in the same manner at $S$.  It can be checked
straightforwardly.
The property $(5.21)$  follows  from the fact that $\psi$ is in essence
a scalar under
changes of coordinates along  the gauge orbits and can be verified by using
Propositions 2 and 3 above.
Finally the property $(5.22)$ reflects  the  fact  that  $\psi$  provides a
representation of the diffeomorphism group, i.e., $\psi_{R\rightarrow S_2} =
(\psi_{R\rightarrow S_1})_{\rightarrow S_2}$.  It can be established by using
Proposition 4. We leave the details of  the calculations to the reader.

{}From $(5.21)$, one gets
$$\eqalign{\Z{a}{a}{0}{1}\ \hG{a}\ \psi &= -i \Z{a}{a}{0}{1}
\ {\cal L}_{\vx{a}}\ \psi\cr
&= -i {\cal L}_{\Z{a}{a}{0}{1}\vx{a}}\psi = 0,\cr}$$
and from $(5.22)$  and $(5.21)$,
$$\eqalign{\left[\hG{a},\hG{b}\right]\ \psi &= - \left[{\cal L}_{\vx{a}},{\cal
L}_{\vx{b}}\right]\ \psi\cr
&= - {\cal L}_{[\vx{a},\vx{b}]}\ \psi\cr
&={\cal L}_{\C{c}{a}{b}\vx{c}}\ \psi\cr
&= \C{c}{a}{b}\ {\cal L}_{\vx{c}}\ \psi = i \C{c}{a}{b} \hG{c}.\cr}$$
This proves $(5.13)$ and $(5.14)$ \footnote{{$^\dagger$}}{Because of $(5.21)$,
one can interpret ${\cal  L}_{\vx{a}}$  as  a  kind  of covariant derivative.
This derivative has zero connection  due  to  $(5.22)$.}.  In addition, the
property $(5.20)$ guarantees the covariance of  the Dirac conditions under
changes  of coordinates and redefinitions of the  constraints  and  of  the
reducibility functions.
\vskip2pc
{\bf 5.4. OBSERVABLES LINEAR IN THE MOMENTA}
\vskip1pc
The quantum definition of an observable $A({\bf q},{\bf  p})$ linear in the
momenta
$$A = a^i({\bf q})\ p_i,\eqno(5.23)$$
$$\{A,\G{a}\} \approx 0,\eqno(5.24)$$
can  be done along the same geometrical lines.   Indeed,  it  follows  from
$(5.23)$ and $(5.24)$ that
$${\cal L}_{\vec\alpha} \vx{b} = - X^{c_0}_{b_0} \vx{c},\eqno(5.25)$$
and thus also
$${\cal L}_{\vec\alpha} \Z{a}{a}{{k-1}}{k} = X^{c_k}_{a_k}
\Z{a}{c}{{k-1}}{k} - X^{a_{k-1}}_{c_{k-1}} \Z{c}{a}{{k-1}}{k},\eqno(5.26)$$
where $\vec\alpha$ as the vector field on ${\cal Q}$ defined by
$$\vec\alpha = a^i({\bf q})\ {\partial\over\partial q^i}.\eqno(5.27)$$
The functions $X^{b_k}_{a_k}({\bf q})$ are subject  to identities similar
to  that fulfilled by $\D{b}{a}{a}{k}$.  By  repeating  the  steps  leading
to $(5.17)$, one gets, for any vector field $\vec\alpha$
$${\cal L}_{\vec\alpha} \psi = a^i \ \psi_{,i} +
{1\over 2} \left(a^i_{,i} - X^{b_0}_{b_0} + \sum^L_{k=1}    (-)^k
X^{b_k}_{b_k}\right)\ \psi.\eqno(5.28)$$
The group property of diffeomorphisms implies
$$\left[{\cal L}_{\vec\alpha},{\cal L}_{\vec\beta}\right]\ \psi = {\cal
L}_{\left[\vec\alpha,\vec\beta\right]}\ \psi,\eqno(5.29)$$
and
$$\eqalign{\left[{\cal L}_{\vec\alpha},{\cal L}_{\vx{a}}\right] \psi &= {\cal
L}_{\left[\vec\alpha,\vx{a}\right]} \psi\cr &= - X^{b_0}_{a_0}
{\cal L}_{\vx{b}}
\psi,\cr}\eqno(5.30)$$
(from $(5.25)$ and $(5.21)$).   Note  that  ${\cal L}_{\mu\vec\alpha} \psi$
$\not=$    $\mu    {\cal   L}_{\vec\alpha}  \psi$  in    general    (unless
$\vec\alpha=\alpha^{a_0} \vx{a}$).

One can take for the quantum operator  $\hat  A$  associated with $A$ minus
$i$ times the Lie derivative along $\vec\alpha$,
$$\hat A \psi = -i {\cal L}_{\vec\alpha}\psi.\eqno(5.31)$$
Because of $(5.30)$, $\hat A$ maps Dirac states  on  Dirac  states  so that
$(5.31)$ is consistent.
\vskip2pc
\vfill
\eject
{\bf 5.5 CONCLUSIONS}
\vskip1pc
In this section, we have shown that the transformation law $(5.12)$ for the
Dirac wave functions leads to a quantization procedure that is  consistent.
The absence of anomaly in the algebra of the quantum constraints  would not
have been achieved had we taken $\psi({\bf q})$ to be merely a  density  of
weight  one-half  in  ${\cal Q}$, without weight for the redefinitions of the
constraints. Indeed for densities of weight 1/2, one does not have
${\cal  L}_{\mu\vx{a}}    \psi=    \mu{\cal    L}_{\vx{a}}    \psi$,  (unless
$\x{i}{a}\partial_i\mu=0$) and thus, $(5.13)$ and $(5.14)$ would fail.
The extra terms in $(5.17)$ containing the structure functions
are therefore crucial.

The  quantization  procedure  allows  also  for  a  geometrical  consistent
definition  of  the  observables  that are linear  in  the  momenta.    The
resulting expression for $\hat A$ is  not  formally Hermitian
in  the  scalar  product  $\int  d{\bf  q} \chi^*({\bf q})  \psi({\bf  q})$.
However this is no harm because $\int d{\bf  q}\chi^*({\bf q}) \psi({\bf q})$
is not the physical scalar product.  We shall derive  the  correct physical
scalar  product  and  prove  its  equivalence  with the reduced phase space
quantization  scalar  product  in the next section.  This requires a better
understanding of the transformation law $(5.12)$.
\vskip4pc
\vfill
\eject
\centerline{\bf    6. EQUIVALENCE OF THE    REDUCED    PHASE SPACE}
\centerline{\bf AND DIRAC QUANTIZATIONS}
\vskip2pc
{\bf 6.1. EQUIVALENCE OF PHYSICAL SPECTRUM}
\vskip1pc
The reduced  configuration  space  ${\cal  Q/G}$  is  the  quotient  of the
configuration space ${\cal  Q}$  by  the  gauge orbits.  There is a natural
map $\pi:$ ${\cal Q}\rightarrow  {\cal Q/G}$ that maps any point of ${\cal
Q}$ on its equivalence class  in  ${\cal Q/G}$.  If the wave functions of the
reduced phase space quantization were scalars  in  ${\cal Q/G}$, they would
induce by pull-back scalars in ${\cal Q}$  that  are  constants  along  the
gauge orbits:
$$f  \in C^\infty({\cal Q/G}) \longrightarrow \pi^* f =  f\circ  \pi  \in
C^\infty({\cal Q}),\ \ \ \ \ \ \ \ \ \ \ \partial_{\vx{a}} (\pi^*f) = 0.$$
However, the wave functions of the reduced phase space  quantization  are
not scalars in ${\cal Q/G}$.  Rather, they are densities of weight 1/2.  We
show  here  that they induce objects on ${\cal Q}$ with the  transformation
law $(5.12)$  which have, furthermore, zero Lie derivative $(5.17)$ along
the gauge generators  $\vx{a}$.

Consider the vectors $\partial/\partial
y^\alpha$ tangent to the  coordinate  lines $(y^\alpha)$ in a local chart
of  ${\cal Q/G}$.  Let  $(\partial/\partial  y^\alpha)^R\equiv
{\vec Y}_\alpha$  be vector  fields  in  ${\cal  Q}$ that
project  down  to  $\partial/\partial y^\alpha$.
At  each point of ${\cal  Q}$,  the  vectors  ${\vec Y}_\alpha$  and
$\vx{a}$ provide an overcomplete set of vectors, which  is a basis if and
only  if the constraints are independent.  Among the  $\vx{a}$,  one  can
choose  locally $m$ independent vectors ${\vec\xi}_A$, which, together with
the $Y_\alpha$,  form  a  basis of the tangent space to ${\cal Q}$.  Let us
expand  the  vectors    $\partial    /\partial    q^i$    in    terms  of
$(Y_\alpha,{\vec\xi}_A)$,
$${\partial \over\partial q^i} = \mu^\alpha_i({\bf q}) {\vec Y}_\alpha +
\mu^A_i({\bf q}) \vec\xi_A,\eqno(6.1)$$
and let us define
$$\mu ({\bf q}) \equiv \left\vert det
(\mu^\alpha_i,\mu^A_i)\right\vert,\eqno(6.2)$$
Even though the vectors ${\vec Y}_\alpha$ are not unique
(${\vec Y}_\alpha \rightarrow
{\vec Y}_\alpha + k^A_\alpha {\vec\xi}_A$), the determinant is unambiguous
since
the ambiguity in ${\vec Y}_\alpha$ simply modifies
the rows $\mu^\alpha_i$ by linear combinations of the rows $\mu^A_i$.
\vskip2pc
{\bf Proposition 5.} The determinant $\mu ({\bf q})$ transforms as

$$\mu'\left({\bf q}'({\bf q})\right)\ = det\ {\partial y'^{\alpha}
\over \partial y^\alpha}\ det\ {\partial q^i
\over     \partial    q'^i}\    \left( det\ A_A\ ^B\right)^{-1}\
\mu({\bf q}).\eqno(6.3)$$
under the transformations
$$q^i \longrightarrow  q'^i = q'^i (q^j)\eqno(6.4)$$
$$y^\alpha \longrightarrow  y'^{\alpha} = y'^{\alpha} (y^\beta)\eqno(6.5)$$
$${\vec\xi}_A \longrightarrow {\vec\xi}'_A = A_A\ ^B\ {\vec\xi}_B,\eqno(6.6)$$
\vskip1pc
{\bf Proof.} This simply follows from standard properties of determinants.
\vskip2pc
{\bf Corollary.}  Let  $\psi({\bf q})$ be a density of weight one-half in
the  reduced  configuration    space,   and  $\pi:$  $q^i  \rightarrow  $
$y^\alpha=y^\alpha({\bf q})$  the  projection  from  ${\cal Q}$ to ${\cal
Q/G}$.
Then,        the        function     $\psi({\bf    q})=\left\vert\mu({\bf
q})\right\vert^{1/2}\ \psi\left(y^\alpha({\bf q})\right)$ is a density of
weight  1/2  in  ${\bf  q}$-space  that transform  with  $\vert  det  A_A\
^B\vert^{-1/2}$ under ${\vec\xi}_A \rightarrow {\vec\xi}'_A = A_A\ ^B
\ {\vec\xi}_B.$
\vskip1pc
{\bf Proof.} Obvious.

\vskip2pc
If    the    vectors   $\vx{a}$  are  independent  (i.e.,  $m_0=m$    and
$\vx{a}\equiv{\vec\xi}_A$)  the  analysis  is  done.    But  if they  are
dependent, this is not the whole story.  In that case,  the dimension $m$
of the tangent space $T_{\cal G}$ to the orbits is given
by eq.$(2.10)$, which can be written as
$$m = m_0 - (m_1 - (m_2 - (.........- m_L)...)),$$
so strongly suggesting that $T_{\cal G}$, should be regarded as a multiple
quotient space.
\medskip
Let us begin by considering the simplest case $L=1$. In this case we
regard $T_{\cal G}$, at each point ${\bf q}\in {\cal Q}$,
as a quotient space $V_0/V_1$, where $dim\ V_0=m_0$ and
$dim\ V_1=m_1$; then $m=m_0-m_1$ in agreement with eq.$(2.10)$. We define the
space  $V_0$ as a vector space generated by  $m_0$  linearly  independent
vectors $\{\vXi{a}\}$,  $a_0=1,...,m_0$.    In  $V_0$,  the $\{\vXi{a}\}$
forms a basis. We demand that in the quotient $V_0/V_1$,
the vectors $\vXi{a}$ are mapped on the vectors $\vx{a}$,
$${\vXi{a}\vert}_{T_{\cal G}} = \vx{a}.\eqno(6.7)$$
This is the case if we take $V_1$ to be the space generated by
the  vectors $\vX1{a}\equiv  \Z{a}{a}{0}{1}\  \vXi{a}$,  $a_1=1,...,m_1$.
Indeed, these vectors are mapped on zero,
$${\vX1{a} \vert}_{T_{\cal G}}
\equiv {\Z{a}{a}{0}{1}\ \vXi{a} \vert}_{T_{\cal G}} =
\Z{a}{a}{0}{1}\ \vx{a} = 0,\eqno(6.8)$$
as they should.
Since the order of reducibility is $L=1$, then eq.$(2.9.b)$ tell us that
$\{\vX1{a}\}$
is a set of $m_1$ linearly independent vectors. So we can replace the basis
$\{\vXi{a}\}$ by $\{\vec\Xi_A,\vX1{a}\}$, where the $\vec\Xi_A$'s are the $m$
vectors associated with the basis $\{\vec\xi_A\}$ of $T_{\cal G}$ via the
eq.$(6.7)$. Therefore it is clear that $T_{\cal G}$
can be regarded as $V_0/V_1$, where the equivalence relation is such that the
vectors $\vX1{a} \in V_1$ are identified with zero. One has
$$\vXi{a} = \mu^A_{a_0}({\bf q})\ \vec\Xi_A + \mu^{a_1}_{a_0}({\bf q})\
\vX1{a}.
\eqno(6.9)$$
Define
$$\mu_0 ({\bf q}) \equiv \left\vert det
(\mu^A_{a_0},\mu^{a_1}_{a_0})\right\vert.\eqno(6.10)$$
Again $\mu_0 ({\bf q})$  does  not  depend  on  how  one  choose  to lift
${\vec\xi}_A$.
\vskip2pc
{\bf Proposition 6.} The determinant $\mu_0 ({\bf q})$ transforms as
$$\mu_0 \longrightarrow \mu'_0 = \left\vert det\ A_A\ ^B\right\vert\
\left\vert det\ A_{a_1}\ ^{b_1}\right\vert\ \left\vert det\ A_{a_0}\ ^{b_0}
\right\vert\ \mu_0,\eqno(6.11)$$
under a redefinition
$${\vec\xi}_A \longrightarrow
{\vec\xi}'_A = A_A\ ^B\ {\vec\xi}_B,\eqno(6.12)$$
$$\vXi{a} \longrightarrow {\vec\Xi}'_{a_0} = A_{a_0}\ ^{b_0}\ \vXi{b},
\eqno(6.13)$$
$$\vX1{a} \longrightarrow {\vec\Xi}'_{a_1} = A_{a_1}\ ^{b_1}\ \vX1{b},
\eqno(6.14)$$
(which are equivalent to the transformation laws $(5.7)$, $(5.10)$,
and $(5.11)$ for $k=1$).
\vskip1pc
{\bf Proof.} Obvious.
\vskip2pc
{\bf Corollary.} The function
$\psi({\bf q}) = \mu^{1/2}\ \mu_0^{-1/2}\ \varphi\left(y^\alpha({\bf
q})\right)$ is  a  density  of  weight  1/2  in  ${\cal  Q}$  that
transforms with $\left\vert det\ A_{a_1}\ ^{b_1}\right\vert ^{1/2}
\left\vert det\ A_{a_0}\  ^{b_0}\right\vert^{-1/2}$ under redefinitions of
the constraints and of  the  $\vX1{a}$.  In particular, it does not depend
on the choice of the intermediate vectors $\vec\Xi_A$.

This corollary is a direct  consequence  of  Propositions 5 and 6, and of
the  fact  that  a  redefinition of  the  constraints  $\G{a}$  yields  a
redefinition of the vectors $\vx{a}$.

For $L=1$ the analysis is done.   If  the  $\Z{a}{a}{0}{1}$ are, however,
not independent, one should keep going and regard  the vector space $V_1$
itself  as  a quotient, etc..., until one reaches the  last  reducibility
stage. For $L > 1$ the set $\{\vX1{a}\}$
is not linearly independent, so no longer we define $V_1$ to be the space
that they generate. Rathe, we denote that subspace of $V_0$ by $W_0$.
Let be $\{\vec\Xi_{A_0}\}$ a basis for $W_0$; then eq.$(6.9)$ now reads,
$$\vXi{a} = \mu^A_{a_0}\ \vec\Xi_A + \mu^{A_0}_{a_0} \vec\Xi_{A_0}.
\eqno(6.15)$$
The relations $(2.9)$ among the $Z$'s mean
$$dim\ W_0 = m_1 - (m_2 - (m_3 - (.......- m_L)...)) = m_0-m,$$
which suggests again to regard $W_0$ as a multiple quotient space. So,
define a
vector space $V_1$, $dim\ V_1=m_1$, with basis
$\{\vec\aleph_{a_1}\}$,  and  consider  the  subspace   $W_1\subset  V_1$
generated by the vectors
$$\vec\aleph_{a_2} \equiv \Z{a}{a}{1}{2}\ \vec\aleph_{a_1}, \ \ \ \ \ \ \
a_2=1,...,m_2.\eqno(6.16)$$
These are not linearly independent if $L>2$. The dimension of $W_1$ is,
according to eq.$(2.9.b)$,
$dim\ W_1=m_2-(m_3-(m_4.....-m_L)...))=m_1-m_0+m$.
Let be $\{\vec\aleph_{A_1}\}$ a basis for  $W_1$.    The  quotient  space
$V_1/W_1$    can  be  identified  with  $W_0$,  and    the    image    of
$\vec\aleph_{a_1}$  in the mapping $V_1\rightarrow W_0$ (which we  denote
by    ${\vec\aleph_{a_1}\vert}_{W_0}$)    can    be    identified    with
$\vec\Xi_{a_1}$ since one has
$${\vec\aleph_{a_2}\vert}_{W_0} = \Z{a}{a}{1}{2}
\ {\vec\aleph_{a_1}\vert}_{W_0} = \Z{a}{a}{1}{2}\ \vec\Xi_{a_1}
= \Z{a}{a}{1}{2}\ \Z{a}{a}{0}{1}\ \vec\Xi_{a_0} = 0.$$
Then a basis for $V_1$ is $\{\vec\aleph_{A_0},\vec\aleph_{A_1}\}$, where
$\vec\aleph_{A_0}$ is any vector  projecting  to  $\vec\Xi_{A_0}$  in the
mapping $V_1\rightarrow W_0$.
The $\vec\aleph_{a_1}$'s can be expanded in this basis:
$$\vec\aleph_{a_1} = \mu^{A_0}_{a_1}\ \vec\aleph_{A_0} + \mu^{A_1}_{a_1}\
\vec\aleph_{A_1}.\eqno(6.17)$$
By using the same argument for $W_1$ and so on, we will obtain
$$T_{\cal G} = V_0/(V_1/(V_2/........V_L)...)).$$
One can define
$$\mu \equiv \left\vert det(\mu^\alpha_i,\mu^A_i)\right\vert,\eqno(6.18.a)$$
$$\mu_0 \equiv \left\vert
det(\mu^A_{a_0},\mu^{A_0}_{a_0})\right\vert,\eqno(6.18.b)$$
$$\mu_1 \equiv \left\vert
det(\mu^{A_0}_{a_1},\mu^{A_1}_{a_1})\right\vert,\eqno(6.18.c)$$
$$.$$
$$.$$
$$.$$
$$\mu_{L-1} \equiv \left\vert
det(\mu^{A_{L-2}}_{a_{L-1}},\mu^{a_L}_{a_{L-1}})\right\vert ,\eqno(6.18.d)$$
\smallskip
where the  $\mu$'s  are the coefficients that appear in the
equations generalizing $(6.17)$.
These determinants do not depend  on  how  one lifts the basis vectors of
$W_k$ to $V_{k+1}$. One has,

{\bf Proposition 7.} Let $\varphi(y^\alpha)$ be a density of weight one-half
on the reduced configuration space ${\cal Q/G}$, and let $\psi({\bf q})$ be
defined through
$$\psi({\bf q}) = \mu^{1/2}\ \mu_0^{-1/2}\
\mu_1^{1/2}.....\mu_{L-1}^{1/2}\ \varphi\left(y^\alpha({\bf q})\right).
\eqno(6.19)$$
Then
$${\cal L}_{\vx{a}} \psi({\bf q}) = 0.\eqno(6.20)$$
\vskip1pc
{\bf Proof.} It is  enough  to  check $(6.20)$ in a particular coordinate
system  and with a particular  choice  of  the  constraints  and  of  the
reducibility functions.  We take the  $q^i$  coordinates  to  split  into
gauge invariant coordinates $y^\alpha$ and pure gauge invariant $Q^A$, as
in Section 4, $q^i\equiv (y^\alpha,Q^A)$.  The constraints  can  then  be
taken to be
$$\G{a} = (G_A, G_{{\bar a}_0}),$$
with  $G_A\equiv  P_A$,  $G_{{\bar  a}_0}\equiv  0$.    Similarly,  the
reducibility functions  can  be  taken to be zero or one $^{17,1}$.  With
that choice, the Lie derivative of $\psi$ reduces to
$${\cal L}_{\vec\xi_{A}} \psi \equiv {\partial \psi\over\partial
Q^A},\eqno(6.21.a)$$
$${\cal L}_{\vx{{\bar a}}} \psi\equiv 0.\eqno(6.21.b)$$
Furthermore,    the    vectors    $\vec\xi_A$,    $\vx{a}$,    $\vX1{a}$,
$\vec\aleph_{a_2}$,  etc.,  can  be  taken    in  such  a  way  that  the
determinants  $\mu$, $\mu_0$, $\mu_1$,...,$\mu_{L-1}$, are all  equal  to
one.  Hence $\psi({\bf q}) =\psi(y^\alpha)$ does  not  depend  on  $Q^A$,
establishing  $(6.20)$    ($\Leftrightarrow$    $(6.21)$).    This  proves
Proposition 7.
\bigskip
Conversely, let  $\psi({\bf  q})$  be  an  object  that  transforms as in
$(5.12)$ and that fulfills $(6.20)$. Then $\psi({\bf q}) |\mu|^{-1/2}
|\mu_0|^{1/2}\ ...\ |\mu_{L-1}|^{(-)^{L+1}/2}$ depends only on $y^\alpha$
and defines a density of weight 1/2 on the reduced configuration space.

We thus conclude that  a  density  of  weight  1/2  in ${\cal Q/G}$ induces
naturally a Dirac state as  defined  in  Section  5  and vice-versa.  The
Dirac and reduced phase-space quantizations give  the  same  spectrum  of
physical states.
\vskip2pc
{\bf 6.2. OBSERVABLES}
\vskip1pc
Similarly, the action of the observables that  are  linear in the momenta
are  equivalent  in  both quantization methods.  Namely,  if  $\psi$  is  a
reduced phase space state and $\psi_D$ the corresponding Dirac state, and
if  $\vec\alpha$ is a vector field in ${\cal Q/G}$ and  ${\vec\alpha}_D$  a
vector field  that  project  down to $\vec\alpha$ (fulfilling accordingly
$[{\vec\alpha}_D,\vx{a}]\sim\vx{b}$),   then  ${({\cal L}_{\vec\alpha}\
\psi)}_D = {\cal L}_{{\vec\alpha}_D}\  \psi_D$.  This simply follows from
the group property of mappings  and  the  commutativity of the following
diagram
$$\eqalign{&\gamma^t_D\cr
{\cal Q}\  \  \  \  \ \ \ \ \ \ &\longrightarrow\ \ \ \ \ \ \ \ \ \ {\cal
Q}\cr
\pi\ \downarrow\ \ \ \ \ \ \ \ \ \ \ &\ \ \ \ \ \ \ \ \ \ \ \ \downarrow\
\pi\cr
&\gamma^t\cr
{\cal Q/G}\  \  \ \ \ \ \ \ \ \ &\longrightarrow\ \ \ \ \ \ \ \ \ \ {\cal
Q/G},\cr}\eqno(6.22)$$
in which $\gamma^t$  and  $\gamma^t_D$  are elements of the one-parameter
group  of  diffeomorphisms generated  respectively  by  $\vec\alpha$  and
${\vec\alpha}_D$.  The equivalence of  the linear observables was already
pointed out by Tuynman $^{9}$ in the irreducible case.
\vskip2pc
{\bf 6.3. SCALAR PRODUCT}
\vskip1pc
Finally, we turn to the scalar  product.   One defines the scalar product
of two Dirac states as
$$\int d{\bf q}\ \psi^*({\bf q})\ \varphi({\bf q})\prod^m_{A=1}
\delta\left(\chi^A({\bf q})\right)\ \left\vert det\{G_A,\chi^B\}\right\vert\
|\mu_0|\ |\mu_1|^{-1}\ |\mu_2|\ ...\ |\mu_{L-1}|^{(-)^{L+1}},\eqno(6.23)$$
where $\chi^A=0$ are gauge conditions and $G_A$  a  subset of irreducible
constraint  functions\footnote{{$^\dagger$}} {If there is no such  subset
that  is globally defined, one must introduce a partition  of  unity  and
generalize  $(6.23)$  in  the standard manner.}.  The expression $(6.23)$
is:   $(i)$  invariant  under redefinitions of the constraints and of the
reducibility functions; this is because one has included the factors
$|\mu_0|\ |\mu_1|^{-1}\ |\mu_2|\ ...$ $|\mu_{L-1}|^{(-)^{L+1}}$ besides the
usual Fadeev-Popov determinant;   and  $(ii)$  invariant under changes of
coordinates.  By choosing the  coordinates  and the constraints as in the
proof of the Proposition 7, one can rewrite $(6.23)$ as
$$\int  \psi^*(y^\alpha)\ \varphi (y^\alpha),\eqno(6.24)$$
thereby proving the equivalence of the  scalar  products in the Dirac and
reduced phase space quantizations.
\vskip4pc
\centerline{\bf 7. BRST QUANTIZATION}
\vskip2pc
{\bf 7.1. BRST CHARGE}
\vskip1pc
In the standard BRST method for quantizing a constrained system, the classical
BRST    generator is  realized  as  an    Hermitian    operator on the Hilbert
space of the functions depending on the original variables $q^i$ and
the ghosts $\e{a}{k}$. We say that the theory is free from BRST anomalies,
if a realization $\hO=\hO^\dagger$ can be found such that the classical
property $(3.4)$ becomes  $$\left[\hO,\hO\right]=0,$$ i.e.,
$$\hO ^2 = 0,\eqno(7.1)$$
(remember that the graded commutator is symmetric for fermionic quantities).
\medskip
As it was already proved in  Section 3,  the BRST generator
is linear in the original momenta and the ghost
momenta, when the constraints are linear in the momenta.
So it has the generic structure
$$\Omega  = \sum^L_{k=-1}  \Omega^{a_k}\  \P{a}{k},\  \  \  \  \  \  \  \
\P{a}{-1}\equiv p_i,\eqno(7.2)$$
with
$${\Omega ^{a_k}}^* = (-)^{k+1} \Omega^{a_k},\eqno(7.3)$$
since $\Omega $ is real.
We will prove that the Hermitian ordering
$$\hO = {1\over 2}\left( \hO_R + \hO_L\right)\eqno(7.4.a)$$
$$\hO_R \equiv \sum_{k=-1}^L \Omega ^{a_k} \hP{a}{k},\eqno(7.4.b)$$
$$\hO_L \equiv \sum_{k=-1}^L \hP{a}{k} \Omega ^{a_k}.\eqno(7.4.c)$$
leads to a theory free from BRST anomalies. In fact,
$$\hO = \hO_R - {i\over 2}\sum_{k=-1}^{L} {\partial^{(left)} \over \partial
\e{a}{k}}\Omega ^{a_k},\eqno(7.5)$$
so that
$$\left[\hO,\hO\right] = \left[\hO_R,\hO_R\right] - i
\left[\sum_{k=-1}^{L}  \Omega ^{a_k}
\hP{a}{k},\sum_{j=-1}^{L} {\partial^{(left)} \over \partial
\e{b}{j}}\Omega ^{b_j}\right].\eqno(7.6)$$
The first term in the right-hand side of $(7.6)$ is zero (it is the
same calculation as in the classical case). Hence one gets
$$\left[\hO,\hO\right] = -i
\sum_{k,j=-1}^{L} \Omega ^{a_k} {\partial^{2\ (left)}
\over \partial\e{a}{k} \partial\e{b}{j}}\Omega
^{b_j}.\eqno(7.7)$$
Now the odd vector field  $\Omega^{a_k}$,  defined  on  the configuration
space of the $q$'s and the  $\eta$'s,  has  vanishing  Lie  bracket  with
itself,
$$\sum_{k,j=-1}^{L} \Omega ^{a_k} {\partial^{(left)}\Omega
^{b_j}\over\partial\e{a}{k}} = 0.\eqno(7.8)$$
This  is just the expression of the  classical  nilpotency  of  $\Omega$,
$\{\Omega,\Omega\}=0$.
\vskip2pc
{\bf Proposition 8.} Let $\Omega^{a_k}$ be an odd vector field that
has  vanishing  Lie  bracket  with itself. Then
$$\sum_{k,j=-1}^{L} \Omega ^{a_k} {\partial^{2\ (left)}
\over \partial\e{a}{k} \partial\e{b}{j}}\Omega ^{b_j} =
0.\eqno(7.9)$$
\vskip1pc
{\bf Proof.} By differentiating $(7.8)$ with respect to $\e{b}{j}$,
$$\eqalign {0 & =
\sum_{k,j=-1}^{L} (-)^j\ {\partial^{(left)}\over\partial
\e{b}{j}}\left(\Omega ^{a_k}\
{\partial^{(left)}\over\partial \e{a}{k}}\Omega
^{b_j}\right)\cr & = \sum_{k,j=-1}^{L} (-)^j\ \left(
{\partial^{(left)}\over\partial \e{b}{j}}\Omega ^{a_k}\ \
{\partial^{(left)}\over\partial \e{a}{k}}\Omega ^{b_j} +
(-)^{k(j+1)}\Omega ^{a_k} {\partial^{2\ (left)} \over
\partial\e{b}{j} \partial\e{a}{k}}\Omega ^{b_j}\right)\cr
& = \sum_{k,j=-1}^{L} (-)^j\ \left( {\partial^{(left)}\over\partial
\e{b}{j}}\Omega ^{a_k}\ \  {\partial^{(left)}\over\partial
\e{a}{k}}\Omega ^{b_j} + (-)^{k(j+1)}\Omega ^{a_k}
{\partial^{2\ (left)} \over \partial\e{a}{k}
\partial\e{b}{j}}\Omega ^{b_j}(-)^{(k+1)(j+1)}\right)\cr
& = \sum_{k,j=-1}^{L} \left((-)^j\ {\partial^{(left)}\over\partial
\e{b}{j}}\Omega ^{a_k}\ \  {\partial^{(left)}\over\partial
\e{a}{k}}\Omega ^{b_j} + \Omega ^{a_k}
\ {\partial^{2\ (left)} \over \partial\e{a}{k}
\partial\e{b}{j}}\Omega ^{b_j}\right).\cr}$$
But it is easy to prove  that  the  first  term in the right hand side of
this equation is zero by itself:
$$\eqalign {\sum_{k,j=-1}^{L}(-)^j
{\partial^{(left)}\over\partial \e{b}{j}}\Omega ^{a_k}\ \ &
{\partial^{(left)}\over\partial \e{a}{k}}\Omega ^{b_j}\cr & =
\sum_{k,j=-1}^{L}(-)^j\ (-)^{j+k+1}
\ {\partial^{(left)}\over\partial \e{a}{k}}\Omega ^{b_j}\ \
{\partial^{(left)}\over\partial \e{b}{j}}\Omega ^{a_k}\cr &=
- \sum_{k,j=-1}^{L} (-)^k
{\partial^{(left)}\over\partial \e{a}{k}}\Omega ^{b_j}\ \
{\partial^{(left)}\over\partial \e{b}{j}}\Omega ^{a_k}\cr &=
- \sum_{k,j=-1}^{L} (-)^j
\ {\partial^{(left)}\over\partial \e{b}{j}}\Omega ^{a_k}\ \
{\partial^{(left)}\over\partial \e{a}{k}}\Omega ^{b_j}.\cr}$$
This  proves Proposition 8, and as a corollary,  the  nilpotency  of  the
quantum $\hO$ (eq.$(7.1)$).
\bigskip
Let  us  now    expand  explicitly  $(7.5)$.    One  finds  (recall  that
$-\D{b}{a}{a}{k}$  is  the coefficient  of  $\e{a}{0}\e{a}{k}\P{b}{k}$  in
$\Omega$, see proof of Proposition 2 in Section 3),
$$\hO    =    {\hat\eta}^{a_0}\   \hG{a}  +    \sum^L_{k=0}    \hO^{a_k}\
\hP{a}{k},\eqno(7.10)$$
{\it with the same operator $\hG{a}$ as in the Dirac quantization method}
(equation  $(5.19)$).    That is, {\it $\hG{a}$  is  the  coefficient  of
$\e{a}{0}$ in the $\eta -{\cal P}$ ordering of  the Hermitian BRST charge
$(7.4)$.}  In  this  view,  the  term  ${\x{i}{a}}_{,i}$  comes from  the
reordering    of    the        original        degrees      of    freedom
($(1/2)(\x{i}{a}p_i+p_i\x{i}{a}$    $=$    $\x{i}{a}p_i    -        (i/2)
{\x{i}{a}}_{,i}$), the term $\C{b}{a}{b}$ comes  from  the  reordering of
the  ghosts  degrees of freedom ($\e{a}{0},\P{a}{0}$),  while  the  terms
$\D{b}{a}{a}{k}$  comes  from  the  reordering of the  ghosts  of  ghosts
($\e{a}{k}.\P{a}{k}$).   {\it Thus each generation of ghosts  contributes
to the Dirac constraint operators $\hG{a}$.}

The consistency of the Dirac quantization scheme -- i.e.,  no  anomaly in
$(5.13)$ and fulfillment of $(5.14)$ -- can also be viewed  as  a  direct
consequence  of  the  absence of BRST anomaly.  Indeed, the nilpotency  of
$(7.1)$ implies straightforwardly  $(5.13)$  and  $(5.14)$
as one can see by examining the first terms of ${\hO}^2$.  One can thus
say that the absence  of  anomaly in the algebra of the Dirac constraints
follows from the inclusion in  $\hG{a}$ of the ghost contribution as well
as of the contribution from the  ghosts of ghosts.  It should be noted in
that  respect  that  one  could have achieved  the  classical  nilpotency
condition $\{\Omega,\Omega\}=0$ without the ghosts of ghosts.   But then,
one would not have found $[\hO,\hO]=0$ quantum mechanically,   since  the
$\D{b}{a}{b}{k}$-terms in $\G{a}$ are essential.

We leave  to the reader to check that similar considerations apply to the
the observables that  are  linear  in  the  momenta,  which,  in the BRST
quantization scheme, fulfill $[\hat A, \hO]=0$, see e.g. Ref.[1].
\vskip2pc
{\bf 7.2. BRST PHYSICAL STATES}
\vskip1pc
In the BRST method,  the  physical  states  are  annihilated  by the BRST
charge,
$$\hO \psi = 0.\eqno(7.11)$$
The link with the Dirac  method as developed in the previous sections,
is  obtained  by demanding, in addition,
that the physical states be annihilated by the ghost momenta
$$\P{a}{k} \psi =0,\ \ \ \ \ \ \ \ k=0,...,L.\eqno(7.12.a)$$
Then the physical wave functions $\psi(q^i,\e{a}{k})$  do  not  depend on
the ghosts,
$$\psi =\psi({\bf q}).\eqno(7.12.b)$$
When $(7.12)$ is inserted in $(7.11)$, one gets
$${\hat\eta}^{a_0}\ \hG{a}\ \psi = 0,\eqno(7.13.a)$$
i.e.,
$$\hG{a}\ \psi = 0,\eqno(7.13.b)$$
{\it which are exactly the Dirac physical  state  conditions.}  Hence the
BRST  physical  states fulfilling $(7.11)$ are exactly the  same  as  the
Dirac physical states.

Since we  have  adopted  an  Hermitian ordering for $\hO$, adapted to the
formal scalar product
$\int dq d\eta\  \psi^*(q,\eta)  \chi(q,\eta)$,  we shall require the BRST
wave functions to transform as superdensities of weight 1/2 under changes
of coordinates in the configuration space of the $q^i$ and the ghosts,
$$q^i, \e{a}{k} \rightarrow q'^i, \eta'^{a_k},\eqno(7.14.a)$$
$$\psi  \rightarrow  \psi'  =  \psi    \left\vert    sdet\  {\partial
(q,\eta)\over \partial (q',\eta')}\right\vert^{1/2}.\eqno(7.14.b)$$
This leaves the integral $\int dq  d\eta\ \psi^* \chi$ invariant.  For the
states $(7.11)$, {\it the rule $(7.14)$ exactly yields the transformation
law $(5.12)$ for the Dirac states.} Indeed,  the redefinition $G'_{a_0}$
$=$  $A_{a_0}\  ^{b_0}  \G{b}$  induces  the  transformation $\e{a}{0}  =
A_{a_0}\   ^{b_0}  \eta'^{b_0}$  of  the  ghosts,  in  order  to    leave
$\G{a}\e{a}{0}$    invariant,    $\G{a}\e{a}{0}  =  G'_{a_0}\eta'^{a_0}$.
Similarly, the redefinitions of the reducibility functions are equivalent
to  a  transformation  of  the  higher  order    ghosts.      Hence,  the
transformation law $(5.12)$ has a very direct explanation in terms of the
BRST quantization.

We again leave it to the reader to check  that  the BRST observables that
are  linear  in  the momenta reproduce correctly the Lie derivative  when
acting on the states $(7.11)$ and $(7.13)$.
\vskip2pc
{\bf 7.3. BRST INNER PRODUCT}
\vskip1pc
To  complete the proof of equivalence of the BRST method with  the  Dirac
method, it remains to discuss the scalar product.

Now, if one computes the integral $\int dq d\eta\  \psi^*(q,\eta)
\chi(q,\eta)$  for the  BRST  physical  states  $(7.11)$,  $(7.12)$,  one
obtains an ill-defined result.    The way out is to introduce a so-called
``non minimal sector" (i.e., further  variables  that  do  not change the
physics) and to regularize $\int dq d\eta\ \psi^*(q,\eta)  \chi(q,\eta)$
by  inserting  the  operator  $\exp{-[{\hat K},\hO]}$  between  $\psi^*$  and
$\chi$, and integrating also over the non  minimal  variables,  with  the
natural measure $dq\ d\eta\ d$(non minimal variables).  Here, ${\hat K}$  is
a ghost minus one operator depending on all the  variables  and  chosen  so
that the integral $\int dq d\eta d({\rm non minimal})\ \psi^*(q,\eta)
\exp{[{\hat K},\hO]} \chi(q,\eta)$ is well defined.
The  operator
$\exp{-[{\hat K},\hO]}$ is (formally) equivalent to the unit operator between
physical  states  because  $\hO$  is  (formally)  Hermitian  with that
natural measure.    When  this  regularization  is  appropriately  carried
through, one finds  that the BRST scalar product coincides with the Dirac
scalar product for the states obeying $(7.11)$ and $(7.12)$.

This result is derived  in  detail  in  Ref.[1]  for the irreducible case
(Chapter 14).  It is  easy  to see, by using the invariance of the scalar
product  under  changes  of  coordinates $q^i,  \e{a}{k}  \rightarrow
q'^i, \eta'^{a_k}$, that the same result applies  to the reducible case
as well.
\vskip4pc
\centerline{\bf 8. CONCLUSIONS}
\vskip2pc
In this paper we have established the equivalence of the reduced phase space,
Dirac and BRST quantization methods for reducible gauge systems described
by constraints linear in the momenta. We have shwon that densities of weight
one-half in the reduced configuration space define densities of weight
one-half in the original configuration space, which have a non trivial weight
under redefinitions of the constraints and of the reducibility functions.
Because of this extra variance, the  Lie derivative of the Dirac wave
functions contains extra terms besides those characteristic of ordinary
density of weight one-half. These terms guarantee the absence of anomalies
in the Dirac quantization scheme, as well as the reducibility of the quantum
constraints.
Finally, we have given a BRST interpretation of the Dirac analysis.
In particular, we have shown that the extra anomaly-cancelling terms in the
quantum constraints could be thought of as arising from the ghosts and the
ghosts of ghosts, which play thus a fundamental role.
\vskip4pc
\centerline{\bf ACKNOWLEDGMENTS}
R.F. was partially supported by a research fellowship from
the European Communities, while M.P.  was supported by  a  grant from the
Universit\'e  Libre  de  Bruxelles.    R.F. and  M.P. wish to thank Professor
R.Balescu and the members of the ``Service de
Physique Statistique, Plasmas et Optique non lin\'eaire"
for their  kind  hospitality.
\vskip4pc
{\parindent0pt
{\bf REFERENCES}

1. M.Henneaux and C.Teitelboim, {\it Quantization of Gauge Systems,}
Princeton University Press, Princeton, 1992.\par
2. K.V.Kucha{\v r}, Phys.Rev.D {\bf 34}, 3031; 3044 (1986).\par
3. M.J.Gotay, J.Math.Phys. {\bf 27}, 2051 (1986).\par
4. A.O.Barvinsky and V.N.Ponomariov, Phys.Lett.B {\bf 167}, 289 (1986).\par
5. B.Kostant and S.Sternberg, Ann.Phys.(N.Y.) {\bf 176}, 49 (1987).\par
6. K.V.Kucha{\v r} and C.G.Torre, J.Math.Phys. {\bf 30}, 1769 (1989).\par
7. D.Mc Mullan and J.Paterson, J.Math.Phys. {\bf 30}, 477; 487 (1989).\par
8. C.Duval, J.Elhadad and G.M.Tuynman, Commun.Math.Phys. {\bf 126}, 535
(1990).\par
9. G.M.Tuynman, J.Math.Phys. {\bf 31}, 83 (1990).\par
10. P.Haji{\v c}ek and K.V.Kucha{\v r}, Phys.Rev.D {\bf 41}, 1091 (1990).\par
11. A.O.Barvinsky, Phys.Lett.B {\bf 241},  (1990).\par
12. R.Loll, Phys.Rev.D {\bf 41}, 3785 (1990).\par
13. G.Kunstatter, Class. Quant. Grav. {\bf 9} (1992).\par
14. A.O.Barvinsky and V.Krykhtin, {\it Dirac and BFV Quantization Methods in
the One-Loop Approximation: Closure of the Quantum Constraint Algebra and the
Conserved Inner Product,} preprint Alberta Thy-7-92, 1992.\par
15. M.Henneaux, Phys.Rep. {\bf 126}, 1 (1985).\par
16. I.A.Batalin and E.S.Fradkin,
Phys.Lett. {\bf 122B}, 157 (1983); Rev.Nuovo Cimento {\bf 9}, 1 (1986).\par
17. J.Fisch, M.Henneaux, J.D.Stasheff and C.Teitelboim, Commun.Math.Phys.
{\bf 120}, 379 (1989).\par}
\end